\newcommand{\be}{\begin{equation}}
\newcommand{\ee}{\end{equation}}

\documentclass{PoS}
\usepackage{graphicx}

\title{Studies of $B$ and $B_s$ Meson Leptonic Decays with NRQCD Bottom and HISQ Light/Strange Quarks}

\ShortTitle{Studies of $B$ and $B_s$ Leptonic Decays with NRQCD and HISQ Quarks}

\author{\speaker{Junko Shigemitsu}, Heechang Na\\
      Physics Department, The Ohio State University, Columbus, OH 43210, USA\\
        E-mail: \email{shige@mps.ohio-state.edu}}

\author{Christine Davies\\
    SUPA, School of Physics \& Astronomy, University of Glasgow, Glasgow, G12 8QQ, UK\\
}
\author{Ron Horgan, Chris Monahan\\
    DAMTP, Cambridge University, Cambridge, CB3 0WA, UK\\
}
\author{Peter Lepage\\
    LEPP, Cornell University, Ithaca, NY 14853, USA\\
}

\abstract{ We present a progress report on new calculations of
 $B$ and $B_s$ meson decay constants employing NRQCD heavy and
 HISQ light valence quarks and using MILC $N_f = 2 + 1$ AsqTad lattices.
 Bare quark masses have been retuned in accord with HPQCD's
 new $r_1$ scale. We find significant reductions in
 discretization effects compared to previous calculations
 with AsqTad light valence quarks. Matching of the NRQCD/HISQ
 heavy-light axial vector current is carried out at one-loop
 order including all relevant dimension 4 current corrections.
}

\FullConference{ The XXIX International Symposium on Lattice Field Theory - Lattice 2011\\
July 10-16, 2011\\
Squaw Valley, Lake Tahoe, California}

\begin{document}

\section{Introduction}
Leptonic decays of charged $B$'s such as $B^+ \rightarrow \tau^+ \nu_\tau$, are 
important processes for Cabbibo-Kobayashi-Maskawa (CKM) and Unitarity Triangle (UT) 
physics.  There is currently some tension between $\epsilon_K$, sin($2\beta$), 
$|V_{ub}|$ and $\cal{B}$$(B^+ \rightarrow \tau^+ \nu_\tau)$ and the $B$ meson 
decay constant $f_B$ plays an important role in such analyses.  For instance 
global fit results for $f_B$ from precision electroweak data
 are being compared with ``direct Standard Model predictions'' 
of the decay constant from lattice QCD \cite{Lunghi1,Lunghi2,Laiho1}.
  Reducing errors in the lattice 
determinations of $f_B$ is, hence, a worthwhile high priority goal.

For the $B_s$ meson there are no 
tree-level leptonic decays in the Standard Model (SM). Nevertheless the decay 
constant $f_{B_s}$ is a useful parameter in many decay and mixing 
rates both within and beyond the SM. Furthermore, lattice QCD determinations of 
$f_{B_s}$ can be achieved with smaller errors than for $f_B$ since no light valence 
quarks are involved here.  This combined with the fact that the 
ratio $\frac{f_{B_s}}{f_B}$ is also known more accurately (due to cancellation 
of statistical and many systematic errors) than either of the decay conatants 
on their own, opens the possibility of getting at precision $f_B$ via 
precision values for $f_{B_s}$ and $\frac{f_{B_s}}{f_B}$.  

The HPQCD collaboration has initiated new calculations of the $B$ and 
$B_s$ meson decay constants based on NRQCD bottom  and HISQ 
light and strange valence quarks and employing MILC $N_f = 2+1$ 
configurations.  This improves on previous determinations of 
$f_B$, $f_{B_s}$ and $\frac{f_{B_s}}{f_B}$ by HPQCD that used AsqTad light and 
strange valence quarks \cite{Gamiz}. Simulation details are summarized in Table 1.

\begin{table}
\begin{center}
\begin{tabular}{|c|c|c|c|c|c|c|c|}
\hline
Set &  $r_1/a$ & $m_l/m_s$  & $a m_{l,s}$ &  $aM_b$ &  $N_{conf}$&
 $N_{tsrc}$ & $L^3 \times N_t$ \\
  & & (sea) & (valence) & (valence) &&& \\
\hline
\hline
C1  & 2.647 & 0.005/0.050 & 0.0070 & 2.650  & 1200  &  2 & $24^3 \times 64$ \\
    &&&        0.0489  &  & 1200  & 2 &                  \\
\hline
C2  & 2.618 & 0.010/0.050 & 0.0123  & 2.688 & 1200   & 2 & $20^3 \times 64$ \\
    &&&        0.0492  & &  1200 & 2 & \\
\hline
C3  & 2.644 & 0.020/0.050 & 0.0246 & 2.650 &  600  & 2 & $20^3 \times 64$ \\
   &&&         0.0491  & &  600 &  2 & \\
\hline
\hline
F1  & 3.699 & 0.0062/0.031 & 0.00674 & 1.832 & 1200  & 4  & $28^3 \times 96$ \\
    &&&        0.0337  & & 1200 & 4  &  \\
\hline
F2  & 3.712 & 0.0124/0.031 & 0.0135 & 1.826 & 600  & 4 & $28^3 \times 96$ \\
    &&&        0.0336  & &  600 &  4 & \\
\hline
\hline
 F0  & 3.695  &  0.0031/0.031 & 0.00339 & 1.832 &
 work in & 4 & $40^3 \times 96$ \\
   &&&         0.0339 & & progress & 4  &\\
\hline
\end{tabular}
\end{center}
\caption{
Simulations details on three ``coarse'' and three ``fine'' MILC ensembles.
}
\end{table}

\section{Tuning of Quark Masses}
We use the static potential quantity $r_1 = 0.3133(23)$fm \cite{Davies1}
 to set 
absolute scales and  $r_1/a$ from MILC \cite{Bazavov} 
 for relative scales between 
different MILC ensembles. To fix the bare $b$-quark mass in lattice 
units $aM_b$ we use the spin averaged $\Upsilon$ mass. One calculates,
\be
\overline{M}_{b \overline{b}}
 \equiv \frac{1}{4} \left [ 3 M_{kin}(^3S_1) + M_{kin}(^1S_0) \right ],
\ee
with 
\be
M_{kin} = \frac{p^2 - \Delta E_p^2}{2 \Delta E_p},
\qquad \qquad \Delta E_p = E(p) - E(0),
\ee
and compares with the experimental value
 (adjusted 
for the absence of electromagnetic, annihilation and sea charm quark 
effects in our simulations) of 9.450(4)GeV \cite{Gregory}.  
Results from this tuning are shown in Fig.1.   Errors in the data points 
include statistical and $r_1/a$ errors.  One sees that these are much 
smaller than the 0.7\% error in the absolute physical value of $r_1$.  
To achieve small statistical errors in $M_{kin}$ it was crucial to 
employ random wall sources for the NRQCD $b$-quark propagators.  With 
 point sources errors would have been about 4$\sim$5 times larger.
The $s$-quark mass was tuned to the (fictitious) $\eta_s$ mass of 
0.6858(40)GeV \cite{Davies1}. Having fixed the bottom and strange quark masses on each 
ensemble one can check how well 
$M_{B_s} - \overline{M}_{b \overline{b}}/2$ is reproduced
 in the continuum limit.  The leading dependence on the heavy quark 
mass cancels in this difference, so one is testing how well the lattice 
actions are simulating QCD boundstate dynamics.  Results for this mass 
difference are shown in Fig.2.  Within the $r_1$ scale error and additional 
$\sim$10MeV uncertainty from relativistic corrections to 
$\overline{M}_{b \overline{b}}$ one sees agreement with experiment
 in the continuum limit.

\begin{figure}
\begin{center}
\includegraphics*[height=10.cm,angle=270]{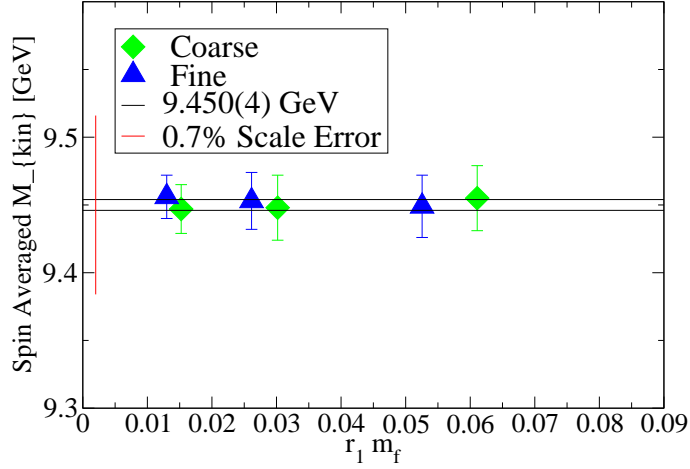}
\caption{
Tuning of the $b$-quark mass using the spin averaged $\Upsilon$ mass.
}
\end{center}
\end{figure}

\begin{figure}
\begin{center}
\includegraphics*[height=10.cm,angle=270]{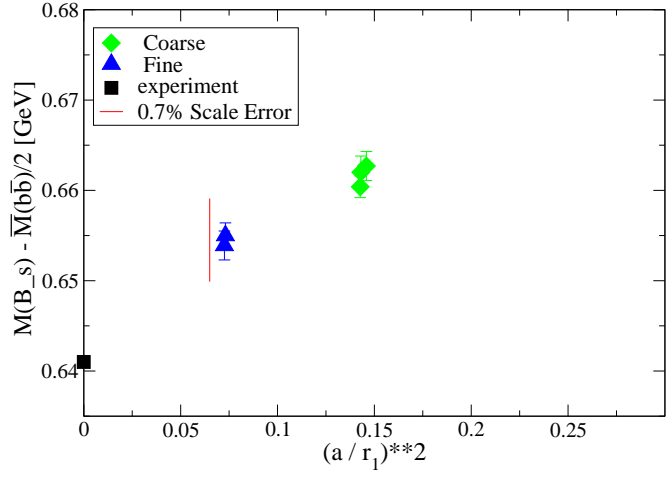}
\caption{
The mass difference $M_{B_s} - \overline{M}_{b \overline{b}}/2$ 
versus the square of the lattice spacing for the first five 
ensembles of Table 1. One sees negligible sea quark mass dependence 
but a noticeable lattice spacing dependence. 
}
\end{center}
\end{figure}

\section{The Currents and Matching}
In the $B_q$ meson rest frame (q = light or strange) the 
decay constant is defined in terms of the temporal component of the  $bq$ heavy-light 
axial vector current $A_0$ as,
\be
 \langle 0 | \; A_0 \; | B_q \rangle_{QCD} = M_{B_q} \; f_{B_q} .
\ee
Simulations are carried out with effective lattice theory currents,
\begin{eqnarray}
 J^{(0)}_{0}(x) & = & \bar q(x) \,\Gamma_0\, Q(x), \\
 J^{(1)}_{0}(x) & = & \frac{-1}{2M_b} \bar q(x)
    \,\Gamma_0\,\mbox{\boldmath$\gamma\!\cdot\!\nabla$} \, Q(x), \\
 J^{(2)}_{0}(x) & = & \frac{-1}{2M_b}  \bar q(x)
    \,\mbox{\boldmath$\gamma\!\cdot\!\overleftarrow{\nabla}$}
    \,\gamma_0\ \Gamma_0\, Q(x),
\end{eqnarray}
and matching through order $\alpha_s, \; \frac{\Lambda_{QCD}}{M}, \;
\frac{\alpha_s}{aM},\; a \alpha_s, \; \alpha_s \frac{\Lambda_{QCD}}{M}$ gives,
\begin{eqnarray}
\langle A_0 \rangle_{QCD}  &=& ( 1 + \alpha_s
 \, 
\rho_0)\,\langle J^{(0)}_0 \rangle + 
 (1 + \alpha_s   \,  \rho_1) \, \langle
J^{(1),sub}_0 \rangle + \alpha_s  \,
 \rho_2 \, \langle J^{(2),sub}_0 \rangle,
   \\
 &&  \nonumber \\
 J^{(i),sub} &=& J^{(i)} - \alpha_s \,   \zeta_{10}
J^{(0)} .
\end{eqnarray}
$\rho_0$, $\rho_1$, $\rho_2$ and $\zeta_{10}$ are the one-loop 
matching coefficients which have recently been calculated for NRQCD/HISQ 
currents.

\section{Preliminary Results and Error Estimates}

Figs.3 \& 4 show our preliminary chiral/continuum extrapolations 
 of $f_{B_s} \sqrt{M_{B_s}}$,
$f_B \sqrt{M_B}$ and 
the ratio $f_{B_s} \sqrt{M_{B_s}}
/f_B \sqrt{M_B}$ based on the first five ensembles of Table 1.  
Work on the sixth ensemble F0, a more chiral fine ensemble, is in progress.  
Extrapolations are carried out using continuum partially quenched 
ChPT for heavy-light decay constants augmented by lattice spacing 
dependent terms. Our preliminary numbers at the physical point are,
\be
\label{results}
f_B = 0.191(9){\rm GeV}, \qquad f_{B_s} = 0.226(10){\rm GeV}, 
\qquad f_{B_s}/f_B = 1.184(19).
\ee

\begin{figure}
\begin{center}
\includegraphics*[height=11.5cm,angle=270]{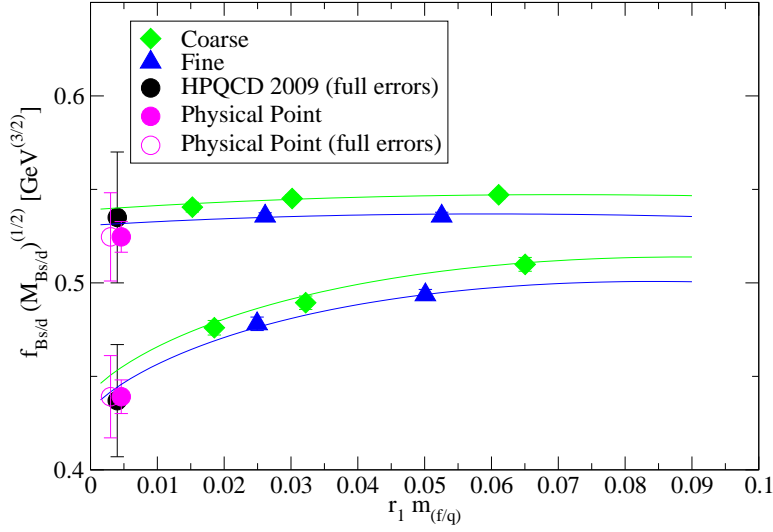}
\caption{
Continuum/chiral extrapolations of $f_{B_s} \sqrt{M_{B_s}}$ (upper curves) and 
$f_B \sqrt{M_B}$ (lower curves). 
The black circles are  
 results at the physical point  from reference \cite{Gamiz} 
using AsqTad light and strange valence quarks.
}
\end{center}
\end{figure}

\begin{figure}
\begin{center}
\includegraphics*[height=11.5cm,angle=270]{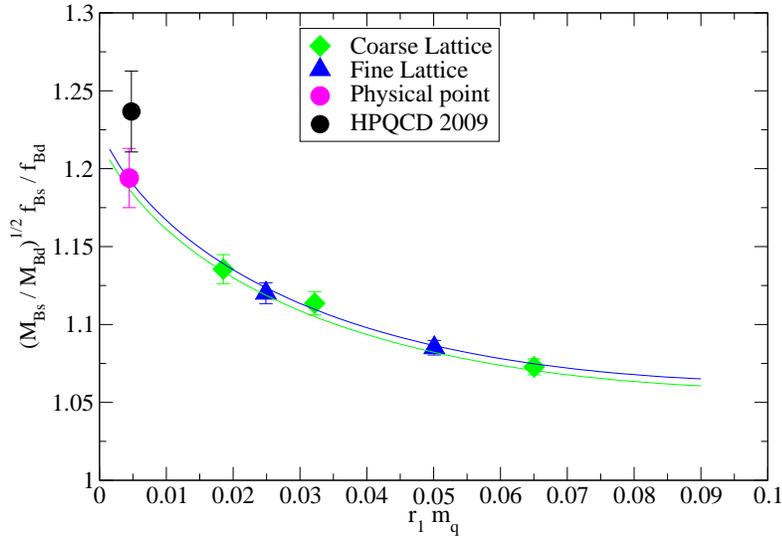}
\caption{
Continuum/chiral extrapolation of the ratio $f_{B_s} \sqrt{M_{B_s}}/
f_B \sqrt{M_B}$.
The black circle is 
the result at the physical point  from reference \cite{Gamiz} 
using AsqTad light and strange valence quarks.
}
\end{center}
\end{figure}

Table 2 shows a preliminary error budget.

The new NRQCD/HISQ numbers are consistent with HPQCD's NRQCD/AsqTad 
results of $f_B = 0.190(13)$GeV, $f_{B_s} = 0.231(15)$GeV, 
$f_{B_s}/f_B = 1.226(26)$ \cite{Gamiz}. 
The reduction in errors in the new calculations 
comes mainly from improvement in discretization errors, smaller $r_1^{3/2}$ 
scale uncertainties and better fitting and extrapolation strategies. 
The striking decrease in lattice spacing dependence as 
one goes from AsqTad to HISQ strange quarks is demonstrated in 
Fig.5, where we compare $f_{B_s} \sqrt{M_{B_s}}$ results on the same 
ensembles for the two different valence quark actions.
 Although this improvement 
in discretization errors is very welcome,  
in both the present and previous calculations the total error is
dominated by the higher order operator matching uncertainty. HPQCD is 
investigating nonperturbative matching strategies for NRQCD/HISQ 
currents which could reduce this error in the future \cite{Koponen}.

In a completely different thrust, we are pursuing an alternate approach to 
 $B$ physics that uses the relativistic HISQ action for heavy quarks 
 with masses $m_H > m_{charm}$ on very 
fine lattices \cite{McNeile,Follana}.
 One can then extrapolate up to the physical $b$-quark staying 
always within a relativistic frame work. A recent result using this method 
gives $f_{B_s} = 0.225(4)$GeV \cite{hisqb}
 in excellent agreement with eq.(\ref{results}), however 
with much reduced errors. Using relativistic heavy quarks enables us to work 
with absolutely normalized currents (based on Ward identities). The main source 
of the larger errors with NRQCD $b$-quarks, namely operator matching uncertainties, 
is thus removed. 
In order to carry out heavy HISQ calculations that can be extrapolated to the 
physical $b$-quark, very fine and hence large lattices are required. Repeating the 
$f_{B_s}$ determination for $f_B$ would be quite expensive.  So for the next couple of 
years we believe the best strategy for precision $B$ physics will be to work with HISQ 
heavy quarks for $B_s$ physics and combine these results with ratios, 
such as $f_{B_s}/f_B$ from NRQCD $b$-quark calculations. The calculations presented 
here are part of this comprehensive approach to precision $B$ physics.

\begin{table}[t]
\begin{center}
\begin{tabular}{|c|c|c|c|}
\hline
 Source  & $f_{B_s}$ &  $f_B$ &  $f_{B_s} / f_B$  \\
        &  (\%)  &  (\%) & (\%)  \\
\hline
\hline
Statistical   &  0.7 &  1.1 & 0.9\\
Scale $r_1^{3/2}$  &  1.1 & 1.1&  ---\\
continuum extrap. &  0.9 & 0.9 &  0.8\\
chiral extrap.  &  0.3 & 1.0&  1.0 \\
$g_{B^*B\pi}$   &  0.1 & 0.1 &   0.1\\
mass tuning    &  0.2 & 0.1 &   0.2\\
relativistic correct.  & 1.0 & 1.0& 0.0\\
operator matching &  4.0 & 4.0&  0.1\\
\hline
 Total  &  4.4 &  4.6&  1.6\\
\hline
\end{tabular}
\end{center}
\caption{
Preliminary error budget}
\end{table}

\begin{figure}
\begin{center}
\includegraphics*[height=11.5cm,angle=270]{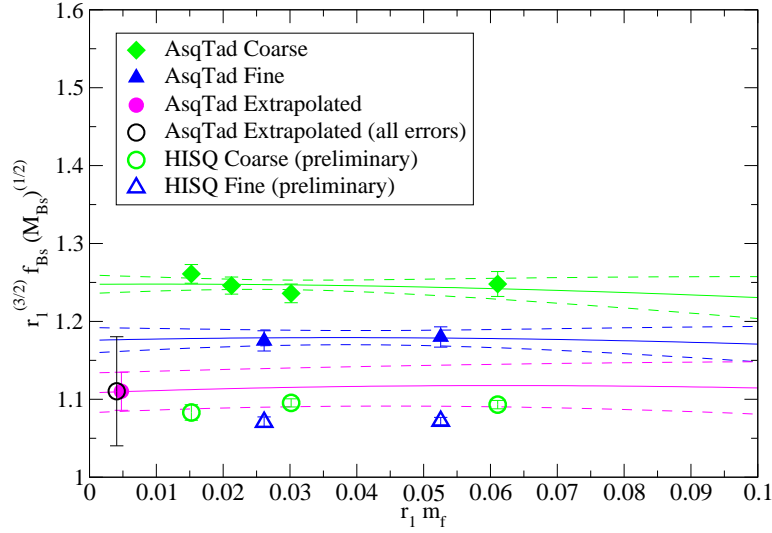}
\caption{
Comparison of lattice spacing dependence between 
NRQCD/HISQ results for $f_{B_s} \sqrt{M_{B_s}}$ from 
the current calculations (open circles and triangles)
and previous results from reference \cite{Gamiz} based on NRQCD/AsqTad quarks.
}
\end{center}
\end{figure}

\noindent
{\bf Acknowledgments}\\
We thank the MILC collaboration for the use of their configurations.  
The numerical simulations were carried out on facilities of the USQCD
Collaboration funded by the Office of Science of the DOE 
and at the Ohio Supercomputer Center.

\end{document}